\documentclass[aps,superscriptaddress]{revtex4}
\usepackage[T1]{fontenc}
\usepackage[utf8]{inputenc}
\usepackage{mathtools}
\usepackage{graphicx}
\usepackage{caption}
\usepackage{subcaption}
\usepackage{xcolor}
\usepackage{soul}
\usepackage{amsmath,amssymb}
\usepackage[colorlinks=true, pdfstartview=FitV, linkcolor=blue, citecolor=blue, urlcolor=magenta, breaklinks=true]{hyperref}
\usepackage{orcidlink}
\usepackage{etoolbox,lineno}%

\begin{document}

\title{Effects of the symmetry energy slope on magnetized neutron stars}

\author{Luiz L. Lopes }

\email{llopes@cefetmg.br}
\affiliation{Centro Federal de Educação Tecnológica de Minas Gerais Campus VIII, CEP 37.022-560, Varginha, Brazil}

\author{César V. Flores}

\email{cezarovf@gmail.com}
\affiliation{Universidade Estadual da Região Tocantina do Maranhão, UEMASUL, Centro de Ciências Exatas, Naturais e Tecnológicas, Imperatriz, CEP 65901-480 Maranhão, Brazil}
\affiliation{
Universidade Federal do Maranhão, Programa de Pós-graduação em Física, 65080-805, São Luís, Maranhão, Brazil}

\author{D\'ebora P. Menezes }

\email{debora.p.m@ufsc.br}
\affiliation{Depto de Física, CFM, Universidade Federal de Santa Catarina, Florianópolis, SC, CP:476, CEP 88.040-900, Brazil}

\begin{abstract}
In this work, we study the effect of the symmetry slope on the observables of weakly and strongly magnetized neutron stars within the chaotic magnetic field approximation. We investigate the impact of the symmetry energy slope in the equation of state, as well as on the observables of neutron stars, by calculating their masses, radii, redshifts, tidal deformabilities, and fundamental-mode gravitational-wave frequencies. We show that the effect of the magnetic field is strong on low mass stars, producing a softer equation of state and correspondingly lower values of radii. Furthermore, the magnetic field also causes a significant drop in the dimensionless tidal parameter even when the effects on the radii are small. At the end of the paper, we discuss the effects of the magnetic field in neutron stars' universal relations.

\end{abstract}

\maketitle

\section{Introduction} \label{introduction}

It is very well known that magnetic fields are of outstanding importance in the study of physical phenomena. Those magnetic fields range from $10^{-11}$ G ( e.g in the human brain) to $10^{18}$ G (e.g in  the early Universe and heavy-ion collisions). Other values include Earth's magnetic fields, which are of the order of $10^{-1}$ G, 50 G present in refrigerator magnets, $10^{5}$ G in modern nuclear magnetic resonance equipment, and $10^{12}-10^{13}$ G in the crust of neutron stars. 

This article focuses on magnetars \cite{Duncan1995, Mereghetti:2015asa}, a special class of neutron stars with surface magnetic fields three orders of magnitude stronger than the ones present in standard neutron stars, i.e., up to $10^{15}$ G at their 
surface. Most of the known magnetars detected so far are isolated objects; i.e., they are not part of a binary system and manifest
themselves as either transient X-ray sources, known as soft-$\gamma$ repeaters, or persistent anomalous X-ray pulsars. At the moment, only thirty magnetars have been identified \cite{Olausen:2013bpa}, but modern telescopes are a promise for more information.
The NICER telescope has already detected two hot spots in the same hemisphere of an ordinary pulsar, suggesting a magnetic field
configuration more complex than perfectly symmetric dipoles. It is important to emphasize that the magnetars detected so far are isolated objects, but there is no reason to believe that binary systems do not exist. 

Any realistic neutron star model must be able to fulfill some experimental and observational constraints coming from both terrestrial nuclear experiments and neutron stars' observations. To describe the interior of a neutron star, we use an extended version of quantum hadrodynamics (QHD)~\cite{Serot_1992} with the L3$\omega\rho$ parametrization~\cite{Lopes2022CTP}. 
This parameterization is capable of fulfilling five phenomenological constraints at the saturation point: the saturation point itself ($n_0$), the incompressibility, $(K)$, the symmetry energy ($S_0)$, the binding energy per nucleon ($B/A$), and the nucleon effective mass at the saturation density $(M^*/M)$, as discussed in two extensive review papers~\cite{Dutra2014,Micaela2017}. A sixth parameter, the symmetry energy slope $(L)$ is left as a free parameter, so that we can investigate how it affects the equation of state (EOS) and the neutron stars' macroscopic properties for both weak and strong magnetized stellar matter. In the present work, the slope $L$ is considered fully independent of the symmetry parameter $S_0$. However, some studies point out a linear correlation between $S_0$ and $L$ as discussed in Ref.~\cite{HOLT2018PLB}. On the other hand, analyzing  neutron skin thickness of heavy nuclei, Ref.~\cite{Wen2010PRC}, the authors indicates that $S_0$ and $L$ are independent, with $L$ lying between $40 ~<L~<76$ MeV.  

 In the high-density regime, the L3$\omega\rho$ model yields a 
quite stiff
equation of state, capable of supporting neutron stars with masses up to $2.30\,M_\odot$. From the astrophysical observation, we check what values of the slope and the influence of the magnetic field can fulfill the constraints related to the  PSR J0740+6620, with a mass of 2.08 $\pm$ 0.07 $M_\odot$ and a radius in the range of 11.41 km $< R <$ 13.69 km~\cite{Riley2021,Miller2021} and the dimensionless tidal parameter of the canonical star, as pointed in ref.~\cite{AbbottPRL}.

Intending to observe the effect of the symmetry slope on the observables of weakly and strongly magnetized neutron stars, we have computed their mass, radii, redshift, tidal deformability, and the gravitational wave frequency of the fundamental mode. At the end of the paper, we also add a small discussion related to the universal relations and compare our findings with those presented in the literature. We expect that future astronomical observables can help in the discrimination of these magnetized compact stars.

\section{Formalism}

To characterize the interactions between protons and neutrons under the influence of a background magnetic field, we use an extended version of the QHD Lagrangian. We also add leptons to account for the zero charge neutrality condition. After this physical considerations we can see that the total Lagrangian in the mean field approximation (MFA) reads~\cite{Serot_1992,IUFSU,Lopes2012BJP,LopesUNIVERSE2025}:

\begin{eqnarray}
\mathcal{L} = \sum_N\bar{\psi}_N \big [\gamma^0(i\partial_0 -g_{N\omega}\omega_0 - \frac{1}{2}g_{N\rho}{\tau_3} {\rho}_0) -\gamma^j(i \partial_j + e_NA_j) - M_N^* \big ]\psi_N \nonumber \\-\frac{1}{2}m_s^2\sigma_0^2    + \frac{1}{2}m_\omega^2\omega_0^2
+ \frac{1}{2}m_\rho^2{\rho}_0^2 - \frac{1}{3}\kappa M_N (g_{Ns}\sigma_0)^3 - \frac{1}{4}\lambda (g_{Ns}\sigma_0)^4\nonumber \\ + \Lambda_{\omega\rho}(g_{N\rho}^2g_{N\omega}^2{\rho_0^2}\omega_0^2)  +
\sum_l \bar{\psi}_l[(\gamma^\mu(i\partial_\mu -e_lA_\mu) - m_l)\psi_l \label{iufsuL},
\end{eqnarray}
where $\psi_N$ and $\psi_l$ are the Dirac fields of the nucleons and leptons, respectively; the $g's$ are the Yukawa coupling constants that simulate the strong interaction; the $\gamma^\mu = (\gamma^0, \gamma^j)$ are the Dirac matrices; $\tau_3$ is the third Pauli matrix; $M_{N}^* = M_N - g_{N\sigma}\sigma_0$ is the nucleon effective mass, with $M_N =939$ MeV; $m_l = (m_e, m_\mu)$ are the masses of the leptons, 0.511 MeV and 105.6 MeV, respectively; $\kappa$ and $\lambda$ are couling constants related to the $\sigma$-meson self interaction, needed to correct the value of the incompressibility~\cite{Boguta}, $\Lambda_{\omega\rho}$ is the coupling constant related to the non-linear interaction between the $\rho$ and $\omega$ mesons, necessary to correct the value of the slope $L$~\cite{IUFSU}, $A^\mu = (A^0, A^j)$ is the eletromagnetic four-potential, and $e_N$ and $e_l$ are the eletric charge of the nucleons and leptons respectively. The electric charge of neutrons ($n$), protons ($p)$, and both leptons ($l$) is 0, $+e$, and $-e$, respectively. 

By applying the Euler-Lagrange equations, we obtain the equations of motions and the nucleons and leptons energy eigenvalues, which at 
zero temperature are also their chemical potentials. For a constant magnetic field in the $z$ direction, we have~\cite{Peng2007}:

\begin{eqnarray}
 E_n = \sqrt{M^{*}_N +k^2} + g_{N\omega}\omega_0 - \frac{1}{2}g_{N\rho}\rho_0,  \nonumber \\   
 E_p = \sqrt{M^{*}_N + k_z^2 + 2\nu|e|B} + g_{N\omega}\omega_0 + \frac{1}{2}g_{N\rho}\rho_0, \nonumber \\
 E_l = \sqrt{m_l + k_z^2 + 2\nu|e|B}, \label{energyEV}
\end{eqnarray}
where  $n$ referes to neutrons, $p$ to protons,
$\nu$ is a discrete parameter related to the Landau level (LL) and $B$ is the magnetic field. 
The number densities of neutrons, protons, and leptons are:

\begin{eqnarray}
 n_n = \frac{k_{Fn}^3}{3\pi^2},  \quad 
 n_p =  \frac{|e|B}{2\pi^2}\sum_\nu \eta(\nu)k_{zFp}, \nonumber \\
 n_l =  \frac{|e|B}{2\pi^2}\sum_\nu \eta(\nu)k_{zFl},
\end{eqnarray}
where the subscript $F$ indicates Fermi momentum. The $\eta(\nu)$ is the degeneracy of the Landau level $\nu$, wich reads $\eta(\nu) = 1$ for $\nu = 0$ and $\eta(\nu) = 2$ for $\nu > 0$. The summation in $\nu$ in the above expressions runs up to $\nu_{max}$, the largest integer value of $\nu$ for which the squared Fermi momenta of the particle is 
positive and it is defined below for protons and leptons:

\begin{equation}
  \nu_{max(p)} = \frac{\mu_p^2 - M^{*2}_N}{2|e|B} , \quad \mbox{and} \quad  \nu_{max(l)} = \frac{\mu_l^2 - m^{*2}_l}{2|e|B}.
\end{equation}

 To obtain stellar matter EOS, charge neutrality and chemical equilibrium conditions have to be enforced. The equation of state for baryons, leptons, and mesons is derived from statistical mechanics, as discussed in Refs.~\cite{Boguta,Lopes2012BJP,LopesUNIVERSE2025, Prakash_Lattimer_2000,Peng2007,Rabhi_2009} and the references therein.

\subsection{The Chaotic Magnetic Field}

Now, since the magnetic field itself presents energy and pressure, we must explicitly take it into account in the EOS. However, while the energy density of the magnetic field is easily added, the contribution to the pressure is more subtle, due to the possible existence of anisotropies. One way to overcome such difficulties is to use the concept of the chaotic magnetic field, as originally introduced by Zeldovich in the 1960s: {\it  “It is possible to describe the effect of the magnetic field by using the pressure concept only when we are dealing
with a small-scale chaotic field (pag. 158)”}~\cite{Zeldovich}. So, we relax the
condition of a uniform magnetic field in the $z$ direction and assume the chaotic magnetic field, whose stress tensor reads: $diag(B^2/6, B^2/6, B^2/6)$, thus avoiding the anisotropy problem and yielding $p= \epsilon/3$, a radiation pressure formalism. Within the chaotic magnetic field formalism, the EOS reads~\cite{Lopes2015JCAP,Flores2020, Lopes2020EPJA,Wu2017CPC,mateus2022EPJA, Lawerence2025IJMPD}:

\begin{equation}
 \epsilon = \epsilon_M + \frac{B^2}{2}, \quad \mbox{and} \quad p = p_M +\frac{B^2}{6},   
\end{equation}
where $M$ stands for matter. 

Concerning the strength of the magnetic field, powerful magnetars can have a magnetic field around $10^{15}$ G at their surface~\cite{Duncan1995,Duncan1996}. However, due to the scalar Virial theorem~\cite{Shapirobook}, the magnetic field at the neutron stars' core can reach values over $10^{18}$ G. To simulate the magnetic field growing towards the core, we use an
energy density-dependent approach, as done in refs.~\cite{Lopes2015JCAP,Lawerence2025IJMPD}:

\begin{equation}
 B =  B_0 \bigg (\frac{\epsilon_M}{\epsilon_0} \bigg )^\gamma + B_{surface}.   
\end{equation}
where $\epsilon_0$ is the central density of the maximally massive star without the magnetic field, and $\gamma$ is a parameter that controls the growth of the magnetic field. As pointed out in ref.~\cite{Lopes2015JCAP,Lopes2020EPJA,Lawerence2025IJMPD}, for $\gamma~>2$ the results are independent of $\gamma$.

In this work, we use $B_{surface} = 10^{12}$ G and $B_0 = 10^{15}$ G for  "normal", weakly magnetized neutron stars and $B_{surface} = 10^{15}$ G and $B_0 = 3.1 \times 10^{18}$ G for strongly magnetized neutron stars. As pointed out in ref.~\cite{Flores2020}, values of $B_0$ lower than $10^{17}$ G do not affect the main properties of neutron stars. 

\subsection{Validity and limitations of the chaotic magnetic field approximations}

It is worth pointing out that the use of the chaotic magnetic field is a simplification, where the effects of anisotropies are neglected, and the use of the TOV equations is a valid approximation. There is no doubt that the ideal situation is
to use more realistic codes, such as the LORENE, which performs a numerical computation of the neutron star by taking into account the Einstein-Maxwell equations and equilibrium self-consistently. Unfortunately, this calculation is not always feasible for all purposes, besides the very high computational cost inherent to it. LORENE code predicts the existence of a poloidal magnetic field, and stellar deformation is expected for fields above 5$\times 10^{17}$ G.~\cite{Gomes_2017AJ}. Perhaps the main flaw of the chaotic magnetic field is the constant spherical symmetry, even for larger values of the magnetic fields. There are, nevertheless, some desirable features satisfied by the chaotic magnetic field approximation. Using a full-relativistic numerical calculation, ref.~\cite{Chartejee2019PRC} shows that the realistic poloidal magnetic field can be expressed as a multipolar expansion that accounts for the monopole, dipole, quadrupole, and higher-order terms.  
The chaotic magnetic field formalism is a
monopole approximation for the magnetic field profile and, as shown in Fig 3 of Ref.~\cite{Chartejee2019PRC}, the monopole term is dominant in almost the entire star. 
More than that, the monopole term is especially dominant in the neutron star core; when
the magnetic field is stronger. So, in the limit of a very strong magnetic
field, when its influence is larger, our results are very close
to those obtained in Ref.~\cite{Chartejee2019PRC}.
Moreover, one of the main problems of using TOV equations in the presence of strong magnetic fields is the possible appearance of anisotropies in
the momentum-energy tensor. As pointed out in Ref.~\cite{Chartejee2019PRC}
in most cases, an unphysical result $T_{\theta\theta} \neq T_{rr}$ at $r =0$ appears. However, exactly due to the monopole nature of the chaotic magnetic field, we always
obtain $T_{\theta\theta} = T_{rr}$, which guarantees that the TOV equations remain reliable.

\section{Equation of State of magnetized matter}

Here, we follow ref.~\cite{lopescesar} and use the L3$\omega\rho$ parametrization to fix all the coupling constants of the Lagrangian, except $(g_{N\rho}/m_\rho)^2$ and $\Lambda_{\omega\rho}$, which are fixed to reproduce different values of the slope. As pointed out in ref.~\cite{Lopes2022CTP}, the L3$\omega\rho$ parametrization satisfies the five 
most accepted phenomenological 
constraints taken from nuclear physics.  The interested reader can obtain  detailed information on the nuclear matter property constraints in \cite{Dutra2014}.
The parameters of the model, the calculated physical quantities, and the respective constraints are presented in Tab.~\ref{T1}. In the same context, different values of the slope $L$ obtained by varying $(g_{N\rho}/m_\rho)^2$ and $\Lambda_{\omega\rho}$ are presented in Tab.~\ref{T2}.

\begin{table}[h]
\begin{center}
\begin{tabular}{cc|cccc}
\toprule
  & Parameters & &  Constraints  & This model  \\
\toprule
 $\left(g_{\sigma}/{m_s}\right)^2$ & $12.108 \, \mathrm{fm}^2$ &$n_0 (\mathrm{fm}^{-3})$ & 0.148 - 0.170 & 0.156 \\
$\left(g_{\omega}/{m_v}\right)^2$  & $7.132 \, \mathrm{fm}^2$ & $M^{*}/M$ & 0.6 - 0.8 & 0.69  \\
  $\kappa$ & 0.004138 & $K \mathrm{(MeV)}$ & 220.0 - 260.0             &  256.0  \\
$\lambda$ &  - 0.00390  & $S_0 \mathrm{(MeV)}$  & 30.0 - 35.0 &  31.7  \\
- - &  - - & $B/A \mathrm{(MeV)}$  & 15.8 - 16.5  & 16.2 \\
\toprule
\end{tabular}
\caption{Model parameters used in this study and their predictions for symmetric nuclear matter at saturation density. The parametrization was taken from ref.~\cite{lopescesar}, and the phenomenological constraints were taken from refs.~\cite{Dutra2014, Micaela2017}.
}
\label{T1}
\end{center}
\end{table}

\begin{center}
\begin{table}[h]
\begin{center}
\begin{tabular}{cccc}
\toprule
  $L$ (MeV)  & $(g_\rho/m_\rho)^2$ ($\mathrm{fm}^2$)   & $\Lambda_{\omega\rho}$  ($\mathrm{fm}^2$)\\
\toprule
 44.0 &8.40  & 0.0515 \\
 60.0 & 6.16 &  0.0344        \\
76.0 & 4.90 & 0.0171    \\
92.0  & 4.06  & 0   \\
\toprule
\end{tabular}
\caption{Model parameters selected to set the symmetry energy at $S_0$ = 31.7 MeV taken from ref.~\cite{lopescesar}.} 
\label{T2}
\end{center}
\end{table}
\end{center}

\begin{figure*}[h]
\begin{tabular}{ccc}
\centering 
\includegraphics[scale=.58, angle=270]{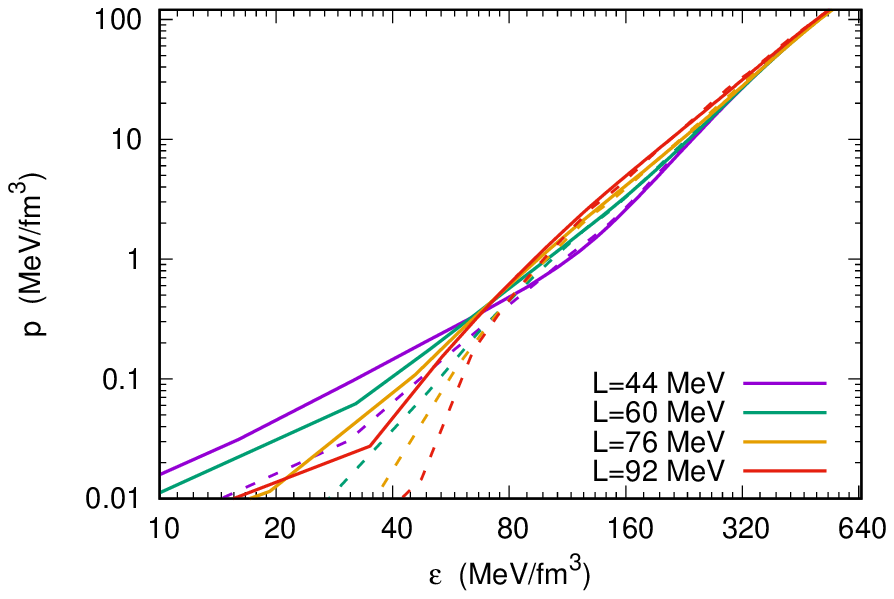} &
\includegraphics[scale=.58, angle=270]{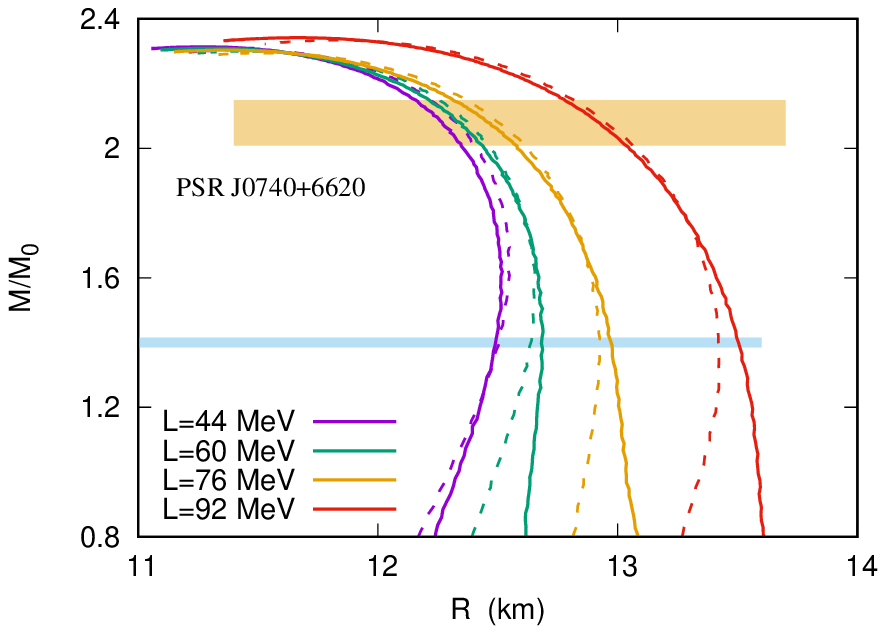} \\
\end{tabular}
\caption{\textbf{(left)} EOS and (\textbf{right}) mass-radius relation for magnetized matter. Solid lines indicate weakly-magnetized neutron stars, and dotted lines indicate strongly-magnetized ones.\label{F1}} 
\end{figure*}

We begin our discussion with the EOS and the corresponding TOV solutions. As already pointed out in the literature, 
the magnetic field~\cite{Lopes2015JCAP} and the slope~\cite{lopescesar} affect low-mass stars more than massive ones.

Using the chaotic magnetic field approximation, we display in the left panel of Fig.~\ref{F1}, the EOS for different values of $L$ with a strong (dashed lines) and a weak (solid lines) magnetic field. The effect of the slope and the magnetic field is barely visible on the EOS in a linear scale. Therefore, we display the EOS on a logarithmic scale. We can notice that by changing $L$, we obtain softer EOS at very low densities ($\epsilon <$ 80 MeV/fm$^3$) for larger values of $L$. At low-to-moderate density, this effect is inverted, and at high densities, all EOSs become almost degenerate, as discussed in ref.~\cite{lopescesar}. The same principle applies to the magnetic field. The magnetic field makes the EOS significantly soft at low densities. At larger densities, the EOSs become slightly stiffer due to the magnetic field. The important feature here is that the magnetic field appears to affect more EOSs with higher values of $L$. 

\section{Stellar Structure}

Since no anisotropies are present, the chaotic magnetic field allows us to use the standard Tolman-Oppenheimer-Volkoff (TOV) equations~\cite{TOV}, which represent the hydrostatic equilibrium of the star. In the following lines we will give an explanation of the physical considerations to obtain these equations.

Assuming that the unperturbed compact star is totally composed of a perfect fluid,  the stress-energy momentum tensor can be expressed as
\begin{equation}
{T}_{\mu\nu} =(\epsilon + p) u_{\mu}u_{\nu} + p{g}_{\mu\nu},
\end{equation}
and the generic background space-time of a static spherical star is expressed through the line element
\begin{equation}
\label{dsz_tov}
ds^{2}=-e^{ \nu(r)} dt^{2} + e^{ \lambda(r)} dr^{2} + r^{2}(d\theta^{2}+\sin^{2}{\theta}d\phi^{2}) .
\end{equation}
where $t, r, \theta, \phi$ are the set of Schwarzschild-like coordinates, and the metric potentials $\nu(r)$ and $\lambda(r)$  are functions of the radial coordinate $r$ only.

The Einstein equations in such a spacetime lead to the following set of stellar structure
equations (Tolman-Oppenheimer-Volkoff equations)

\begin{eqnarray}
\label{tov1}
\frac{dp}{dr} &=& - \frac{\epsilon(p) m}{r^2}\bigg(1 + \frac{p}{\epsilon(p)}\bigg)
	\bigg(1 + \frac{4\pi p r^3}{m}\bigg)\left(1 - \frac{2m}{r}\right)^{-1},  \\
\label{tov2}
\frac{d\nu}{dr} &=& - \frac{2}{\epsilon(p)} \frac{dp}{dr} \bigg(1 + \frac{p}{\epsilon(p)}\bigg)^{-1}, \\
\label{tov3}
\frac{dm}{dr}& =& 4 \pi r^2 \epsilon(p),
\end{eqnarray}
To close this system we need a relation between the pressure and energy density, which is given by the equation of state $\epsilon(p)$. 

Therefore, given an EOS, and by integrating these equations, we can obtain the behavior of the following functions inside the star: $m(r),p(r), \epsilon(r)$ and $\nu(r)$. There is an additional condition in order to consistently determine the $\nu(r)$ function inside the star, it read as 
\begin{equation}
\label{BoundaryConditionMetricFunction}
    \nu(r=R)= \ln \bigg( 1-\frac{2M}{R} \bigg),
\end{equation}
where $R$ is the radius of the star and $M$ its mass.  We can determine the stellar structure, once the previously commented functions have been calculated

In the right panel of Fig.~\ref{F1}, we display the stellar masses M vs he radii R, for different values of $L$ with (dashed lines) and without (solid lines) a magnetic field. In all cases, we use the BPS+BPP~\cite{BPS,BBP} EOS to simulate the neutron star crust. The crust BPS+BBP EOS is used up to 0.01 fm$^{-3}$ for all values of $L$. The core EOS begins when $p_{core} = p_{crust}$, an approach suggested in Gledenning's book~\cite{Glenbook} and widely used. Such an approach implies that the core EOS begins around 0.03 fm$^{-3}$ for $L$ = 44 MeV and slowly grows with $L$, reaching slightly below  0.05 fm$^{-3}$ for $L$ = 92 MeV for the weak magnetic field. { Within strong magnetic fields, there is an increase in the density of the onset of the core EOS. This increase depends on the slope and reaches a density range $\Delta n~\approx$ 0.022 fm$^{-3}$ for $L = 44$ MeV and drops to aroud 0.015 fm$^{-3}$ for $L = 92$ MeV, what implies that in the presence of a strong magnetic field field the core EOS begins in the range $0.050$  fm$^{-3}~<n~<0.065$  fm$^{-3}$, growing with $L$. } Ref.~\cite{Fortin2016} compares the BPS+BBP crust EOS with a unified EOS. They show that for the canonical star, there is a variation in the radius of
60 m $<~\Delta R_{1.4}~<$ 150 m. For a radius of 13 km, this implies
an uncertainty of around 1\%. { Larger deviations are expected for the NL3 model, which is a very different model with a very large slope, $L = 118$ MeV and much stiffer EOS.} The present approach was also used in other studies about the influence of $L$ in non-magnetized neutron stars~\cite{lopescesar,lopes2024PRCb,Lopes2024ApJ,LopesUNIVERSE2025}.
{On the other hand, spinodal calculations point to a core-crust phase transition at higher densities~\cite{Chatterjee2019JCAP}, also that the transition density may not be unique.
However, the very same work shows that the uncertainty in the macroscopic quantities is low, justifying our approach to the crust. It is also worth mentioning that at densities around half of the saturation point, the competition between attractive nuclear force and repulsive Coulomb interaction can turn the nuclear matter into a frustrated system, i.e., the system presents more than one low-energy configuration, which
can cause the onset of unusual nuclear shapes with different geometries. This is called the nuclear pasta phase. Such a discussion is beyond the scope of the present work, and the interested reader is referred to Ref~\cite{Lorenz1993PRL,Nikolaiuniverse2022} and the references therein.} 

Altogether with the TOV solutions presented in Fig.~\ref{F1}, we also display two constraints.  Today, the most well-measured massive pulsar is the  PSR J0740+6620, with a mass of 2.08 $\pm$ 0.07 $M_\odot$ and a radius in the range of 11.41 km $< R <$ 13.69 km~\cite{Riley2021,Miller2021}. Any realistic EOS must be able to fulfill this constraint, which is presented as a yellowish hatched area. The second constraint is related to the radius of the canonical 1.4 M$_\odot$ star. Since it is strongly related to the slope~\cite{Rafa2011},  we use here only a moderate-to-weak constraint presented in ref.~\cite{Annala2018PRL}. Using state-of-the-art theoretical results at low and high baryon densities, the authors constrain the radius of the canonical star to $R_{1.4} < 13.6$ km. Such a constraint is presented as a bluish hatched area. As can be seen, in the present study, all results fulfill both constraints.  The use of this moderate-to-weak constraint enables a systematic investigation of the effects of the parameter $L$ over a broad range, spanning from 44 to 92 MeV. There are, nevertheless, strong constraints related to the canonical 1.4 $M_\odot$ star. For instance, in ref.~\cite{Miller2021}, an upper limit of 13.1 km was appointed.

We now analyze how the strong magnetic field affects the main properties of neutron stars with different masses and slopes. From 1.0 M$_\odot$ to 2.0 M$_\odot$, the results for  all parameters discussed in this work are presented in Tab.~\ref{T3}. Weakly-magnetized neutron stars with $B_0 = 10^{15}$ G are referred to as $B{0}$, while strongly magnetized ones, with $B_{0} = 3.1\times10^{18}$ G, are referred to as $B{1}$.

At first, we can notice that as the slope increases, the radii also increase for all masses. Such an effect was already noted in refs.~\cite{Rafa2011,lopescesar,lopes2024PRCb}, and we can see that it is also true in the presence of the chaotic magnetic field. Furthermore, for lower masses, strongly magnetized stars present a lower radius than weakly magnetized ones. As the masses increase, the radii become closer, and at some point, strongly magnetized neutron stars present slightly larger radii than weakly magnetized ones. Such an effect exists for all values of the slope $L$.  The mass where the behavior of the radius is inverted depends on the slope. For $L =44$ MeV, a 1.4 M$_\odot$ strongly magnetized star already has a larger radius than a weakly magnetized one, while for $L = 92$ MeV, such an effect only takes place for stars above 1.8 M$_\odot$.

The effect of a strong magnetic field is 
 more pronounced
for higher values of $L$, as well as for lower masses. Quantitatively, we can notice that for a 1.0 M$_\odot$, the difference in the radius, $\Delta R$, can reach 0.24 km for $L = 92$ MeV, or only 0.06 km for $L = 44$ MeV. Higher mass stars  are affected by a
$\Delta R~<$ 0.1 km for all values of $L$. All relevant values are presented in Tab.~\ref{T3}. As pointed out, both observational constraints are fulfilled by all values of $L$ and $B_0$.

Another relevant physical quantity that can be obtained from the computations of the stellar structure is the gravitational redshift $Z$. The gravitational redshift is defined as~\cite{Lopes2021EPL}:

\begin{equation}
 Z = \bigg ( 1 - \frac{2M}{R} \bigg )^{-1/2} - 1.   
\end{equation}

\begin{figure*}[ht]
\begin{tabular}{ccc}
\centering 
\includegraphics[scale=.58, angle=270]{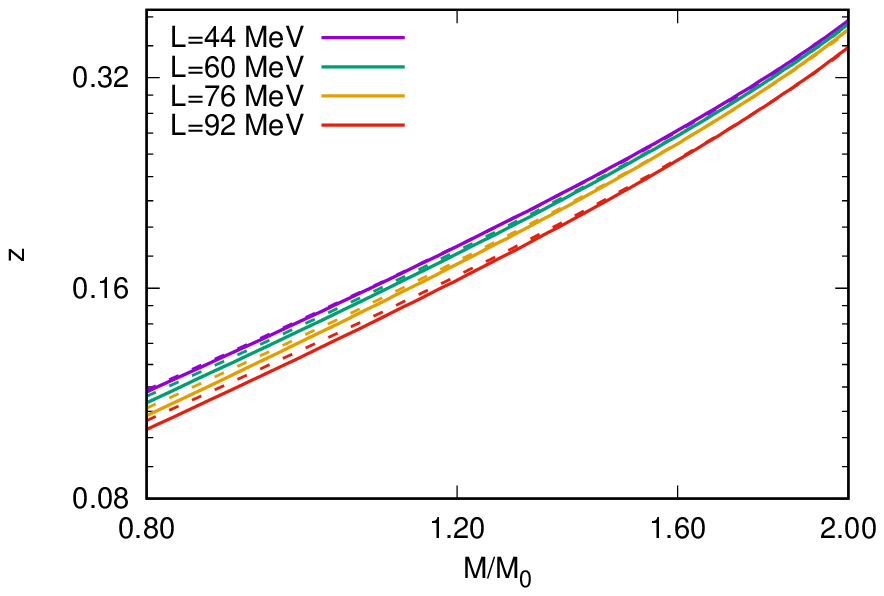} &
\includegraphics[scale=.58, angle=270]{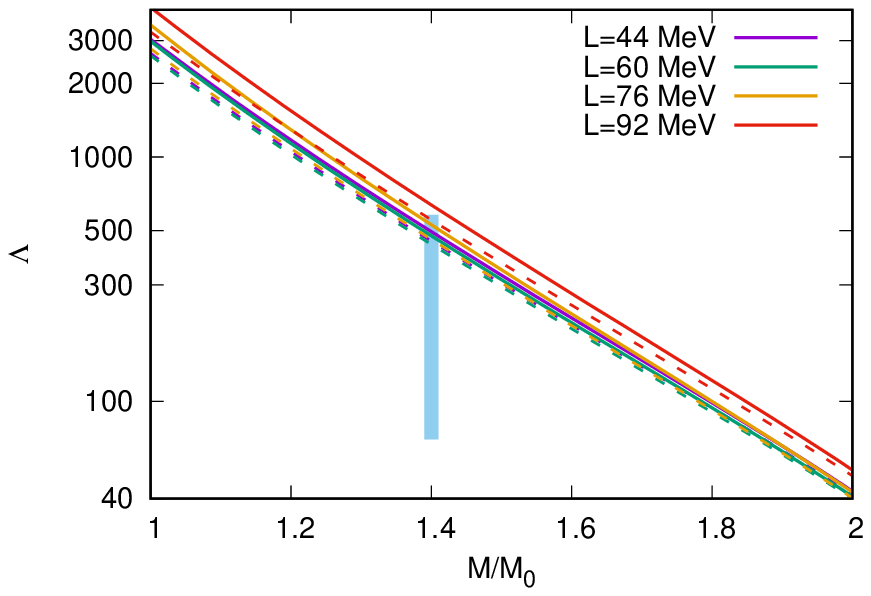} \\
\includegraphics[scale=.58, angle=270]{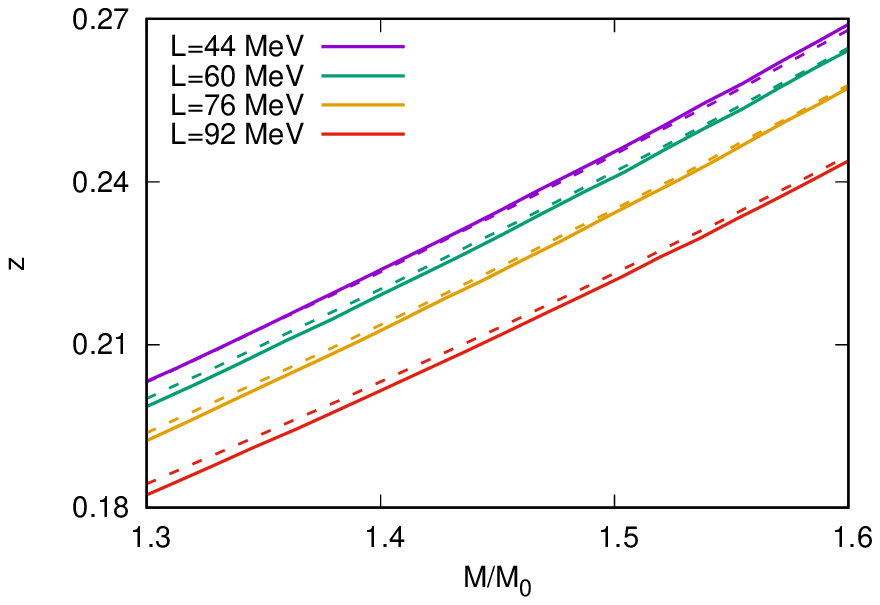} &
\includegraphics[scale=.58, angle=270]{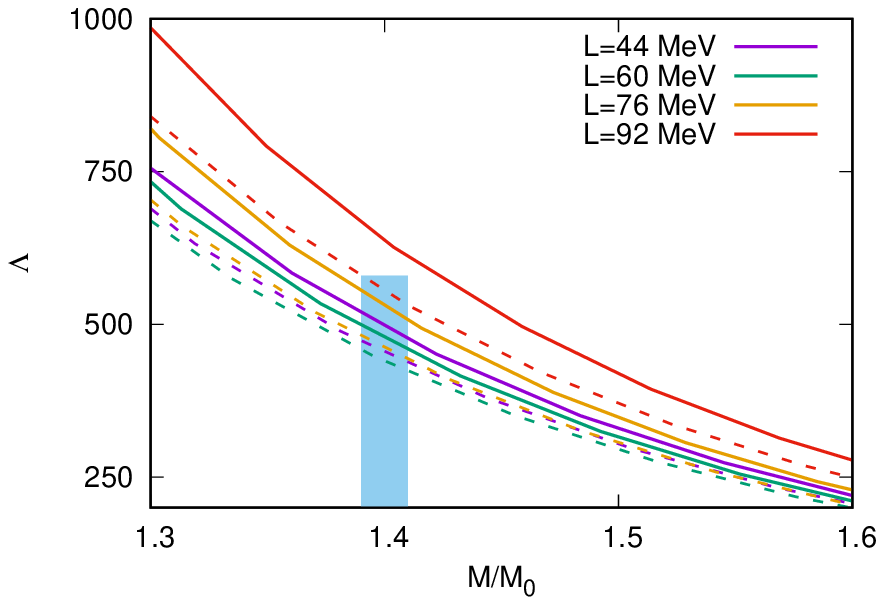} \\
\end{tabular}
\caption{\textbf{(Top-left}) Redshift and (\textbf{Top-right}) dimensionless tidal parameter. \textbf{(Bottom)} Zoom-in focusing on masses between 1.3 and 1.6 solar masses. Solid lines indicate weakly-magnetized neutron stars, and dotted lines indicate strongly-magnetized ones.\label{F2}} 
\end{figure*}

The results for masses in the range 0.4$M_\odot$ $<~M~<$ 2.0$M_\odot$ are displayed on the top-left side of Fig.~\ref{F2}, while a zoomed version is displayed at the bottom-left in the same figure. We can notice that as the mass increases, the redshift also increases, independently of the value of $L$ or $B_0$. We can also see that the redshift is strongly dependent on the slope $L$, but the effects of the magnetic field are small. In the case of the slope, we can see that small slopes can produce values of $Z$ about 0.115 larger than high values of $L$. On the other hand, the presence of a strong magnetic field causes only a $\Delta Z <$ 0.005 in all cases. The values of $Z$ and $\Delta Z$ for some masses are presented in Tab.~\ref{T3}.  Our results indicate that measuring the redshift can be a useful tool for constraining the slope, but not the strength, of the magnetic field.

There are also some constraints related to the redshift. Ref.~\cite{Cottam2020Nature} reports a redshift $z=0.35$ for the EXO0748-676 pulsar. This implies a large mass, varying from 1.89 $M_\odot$ for $L = 44$ MeV up to 1.99 $M_\odot$ for $L = 92$ MeV. On the other hand, Ref.~\cite{Ham2017AeA} 
reports
a redshift $z = 0.205$ for the isolated neutron star RX J0720.4-312, which yields a mass in the range 1.31$M_\odot$ up to 1.41 $M_\odot$ for $L = 44$ MeV and $L = 92$ MeV, respectively.
In 
our analysis, all models satisfy both constraints on $z$. In the near future, simultaneous measurements of the mass and $z$ may provide stringent constraints on the equation of state.

\section{Tidal deformability}

After the computation of the equilibrium configuration we proceed to the study of tidal deformations, which depend on the internal structure of neutron stars and our purpose is to use it as an aditional constraint for the equation of state of magnetars. For this objective we present, in the next lines, some comments about the theory of tidal deformabilities, the main equations and relationships that are necessary for our work. 

The relativistic theory of tidal effects was deduced by Hinderer, Damour and Nagar, Binnington and Poisson \cite{Hinderer_2008,PhysRevD.80.084035,PhysRevD.80.084018}. They concluded that the tidal deformation of a neutron star is characterized by the gravito-electric $K^{el}_{2}$ and gravito-magnetic $K^{mag}_{2}$ Love numbers, where the former is related to the mass quadrupole and the second to the current quadrupole induced by the companion star. Further researches by Flanagan and Hindeler concluded that only a single detection should be sufficient to impose upper limits on $K^{el}_{2}$ at 90\% confidence level
\cite{PhysRevD.77.021502}. Since then intense research has been invested on the computing of Love numbers of neutron stars \cite{PhysRevC.87.015806,PhysRevC.98.065804,PhysRevC.95.015801,PhysRevD.81.123016,Hinderer_2008}.  

In a binary system the induced quadrupole moment $Q_{ij}$ in one neutron star due to the external tidal field ${\cal E}_{ij}$ created by a companion compact object can be written as~\cite{PhysRevD.81.123016,Hinderer_2008},
\begin {equation}
Q_{ij} = -\lambda {\cal E}_{ij},
\end{equation}
where, $\lambda$ is the tidal deformability parameter, which can be expressed in terms of dimensionless $l = 2$ quadrupole tidal Love number $k_2$ as
\begin{equation}
\label{tidal}
\lambda= \frac{2}{3} {k_2}R^{5}.
\end{equation}
To obtain $k_{2}$ we have to solve the following differential equation  
\begin{equation}
r \frac{dy}{dr} + y^2 + y F(r) + r^2Q(r)=0,
\label{ydef}
\end{equation}
where the coefficients  are given by
\begin{equation}
F(r) = [1 - 4\pi r^2(\varepsilon - p)]/E
\end{equation}
and
\begin{align}
Q(r)&=4\pi \left[5\varepsilon + 9p + (\varepsilon + p)\left(\frac{\partial 
p}{\partial\varepsilon}\right)^{-1}-\frac{6}{4\pi r^2}\right]/E 
\nonumber\\ 
&- 4\left[ \frac{m+4\pi r^3 p}{r^2 E} \right]^2,
\end{align}
with $E = 1-2m/r$, $\varepsilon$ and $p$ are the energy density and pressure profiles inside the star. Therefore the Love number $k_2$ can be obtained as
\begin{align}
&k_2 =\frac{8C^5}{5}(1-2C)^2[2+C(y_R-1)-y_R]\times
\nonumber\\
&\times\Big\{2C [6-3y_R+3C(5y_R-8)]
\nonumber\\
&+4C^3[13-11y_R+C(3y_R-2) + 2C^2(1+y_R)]
\nonumber\\
&+3(1-2C^2)[2-y_R+2C(y_R-1)]{\rm ln}(1-2C)\Big\}^{-1},
\label{k2}
\end{align}
where $y_R = y(r = R)$  and $C= M/R$ are the star compactness, $M$ and $R$ are the total mass and radius of the star, respectively.  Equation (\ref{ydef}) has to be solved coupled to the TOV equations. The dimensionless tidal parameter  is defined as:

\begin{eqnarray}
  \Lambda = \frac{2k_2}{3C^5},  
\end{eqnarray}

Adicional discussion can be found in  ref.~\cite{Hinderer_2008,chatziioannou,Chat2020,Flores2020} and the references therein.  
The numerical results are presented on the Top-right side of Fig.~\ref{F2}, while in the bottom-right we present a zoomed version focusing on masses between 1.3 and 1.6$M_\odot$. The observational constraint coming from the GW170817 event detected by LIGO/VIRGO gravitational wave observatories is also presented as a bluish hatched area. It was pointed out in ref.~\cite{AbbottPRL} that the tidal dimensionless parameter of the canonical star 
must lie in the range 70 $<~\Lambda_{1.4}~<$ 580. As shown, this constraint is satisfied for almost all values of $L$ and $B_0$ investigated in the present work. The only exception is $L = 92$ MeV within the weak-magnetic-field limit. 

The presence of the magnetic field always decreases the tidal parameter $\Lambda$. This is valid for all slopes and all masses. Moreover, a strong magnetic field can affect the tidal parameter even more than a change in the slope. For instance, in the case of the canonical star, we see that the lower value of the slope is obtained for $L = 60$ MeV, which within the weak magnetic field, assumes $\Lambda_{1.4} = 501$. However, within the strong magnetic field, values lower than 500 can be obtained for $L =44$ MeV and $L = 76$ MeV as well. This indicates that the tidal parameter is more dependent on the EOS at the low-density limit than the mass-radius relation.

In relation to other mass values, we can see that the effect of the magnetic field is, as expected, stronger for low masses. In the same sense, the effects are larger for larger values of $L$.  Moreover, for massive neutron stars, the dimensionless tidal parameter seems almost independent of both the slope and the magnetic field. 
Regarding these features, we can notice that $\Delta \Lambda$ can reach absolute values above 800 for 1.0 $M_\odot$. This corresponds to variations up to almost $20\%$. We can conclude that for all physical quantities analysed till this moment, the tidal parameter presents the largest sensitivity to the effects of a strong magnetic field. The results for $\Lambda$ and $\Delta \Lambda$ for different masses can be found in Tab.~\ref{T3}.

\section{Neutron Star Oscillations}

The equations governing the nonradial pulsations of a compact star in full general relativity were first investigated by Thorne and Campolattaro~\cite{1967ApJ...149..591T,1970ApJ...159..847C}. They demonstrated that Einstein’s equations for small, nonradial, quasi-periodic oscillations of relativistic stellar models can be reduced to a system of ordinary differential equations for the perturbed variables. In this work, we adopt the formulation developed by Lindblom and Detweiler \cite{1983ApJS...53...73L,1985ApJ...292...12D}, in which Thorne’s perturbation equations are reduced to a system of four ordinary differential equations, allowing the perturbations to be integrated directly in a manner similar to Thorne’s original approach. These equations describe both the fluid oscillations of the star and the associated emission of gravitational waves, which in turn leads to damping of the stellar oscillations.

We assume that the unperturbed, spherically symmetric equilibrium configuration of the compact star is described by a solution of the Tolman–Oppenheimer–Volkoff (TOV) equations and then we consider perturbations in the fluid and metric. For pulsations with spherical-harmonic indices $\ell$ and m and with parity $\pi=(-1)^\ell$, the perturbed metric inside the star, expressed in the Regge–Wheeler gauge~\cite{1957PhRv..108.1063R} takes the form
\begin{eqnarray}
  ds^2 &=& -e^\nu(1+r^\ell H_0^{\ell m}Y_{\ell m}e^{i\omega t})dt^2 +e^\lambda(1-r^\ell H_2^{\ell m}Y_{\ell m}e^{i\omega t})dr^2 \nonumber\\
       & &  -2i\omega r^{\ell+1}H_1^{\ell m}Y_{\ell m}e^{i\omega t}dtdr + r^2(1-r^\ell K^{\ell m}Y_{\ell m}e^{i\omega t})(d\theta^2+\sin^2\theta d\varphi^2),
\end{eqnarray}
where $\omega$ is the frequency, $Y_{\ell m}$ denote the usual scalar spherical harmonics, the functions $e^\nu$ and $e^\lambda$
are the components of the metric of the unperturbed stellar model, while $H_i^{\ell m}(r)$ and $K^{\ell m}(r)$ characterize the metric perturbations. The fluid perturbation is described by the Lagrangian displacement vector $\xi_a$, having components
\begin{eqnarray}
  \xi_r(t,r,\theta,\varphi)        &=& e^{\lambda/2}r^{\ell-1}W^{\ell m}(r)Y_{\ell m}(\theta,\varphi)e^{i\omega t}, \nonumber\\
  \xi_\theta(t,r,\theta,\varphi)   &= &  -r^\ell V^{\ell m}(r)\partial_\theta Y_{\ell m}(\theta,\varphi)e^{i\omega t},\\
  \xi_\varphi(t,r,\theta,\varphi)  &=&   -r^\ell V^{\ell m}(r)\partial_\varphi Y_{\ell m}(\theta,\varphi)e^{i\omega t}.\nonumber
\end{eqnarray}

In the present paper we use the formulation of Lindblom and Detweiler~\cite{1983ApJS...53...73L,1985ApJ...292...12D}, 
consisting of a system of four ordinary differential equations, as given in \cite{Flores_2019}:

\begin{equation}
\frac{d\mathbf{Y}(r)}{dr}=
\mathbf{Q}(r,\ell,\omega)\mathbf{Y}(r)
\end{equation}
for the functions $\mathbf{Y}(r)=(H_1^{\ell m},K^{\ell m},{W}^{\ell m},X^{\ell m})$,
where 
\begin{equation}
X^{\ell m}=-e^{\psi/2}\Delta p^{\ell m}  
\end{equation}
and three algebraic relations, which allow us to compute the remaining functions $\{ H_0^{\ell m},H_2^{\ell m},V^{\ell m}\}$ 
in terms of the others.
We concentrate our attention on normal modes that belong to a particular even parity spherical harmonic $\pi=(-1)^\ell$ 
with the complex frequency  \cite{Flores_2019}
\begin{equation}
 \omega=\sigma+\frac{i}{\tau}.
\end{equation}
The normal modes of the coupled system are defined as those oscillations that lead to purely outgoing waves at spatial infinity. 
The real parts of $\omega$ correspond to the oscillatory frequency
\begin{equation}
    f = Re(\omega)/2\pi = \sigma/2\pi
\end{equation}
and the damping time, 
\begin{equation}
    \tau = 1/Im(\omega),
\end{equation}
which is related to the imaginary part of $\omega$ and corresponds to the radiative energy loss emitted through gravitational waves.

\begin{figure*}[ht]
\begin{tabular}{ccc}
\centering 
\includegraphics[scale=.58, angle=270]{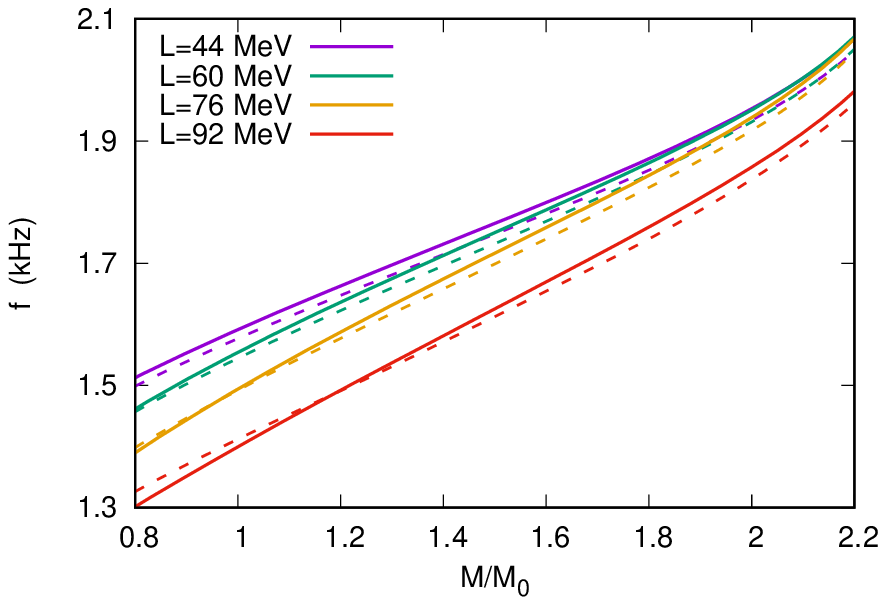} &
\includegraphics[scale=.58, angle=270]{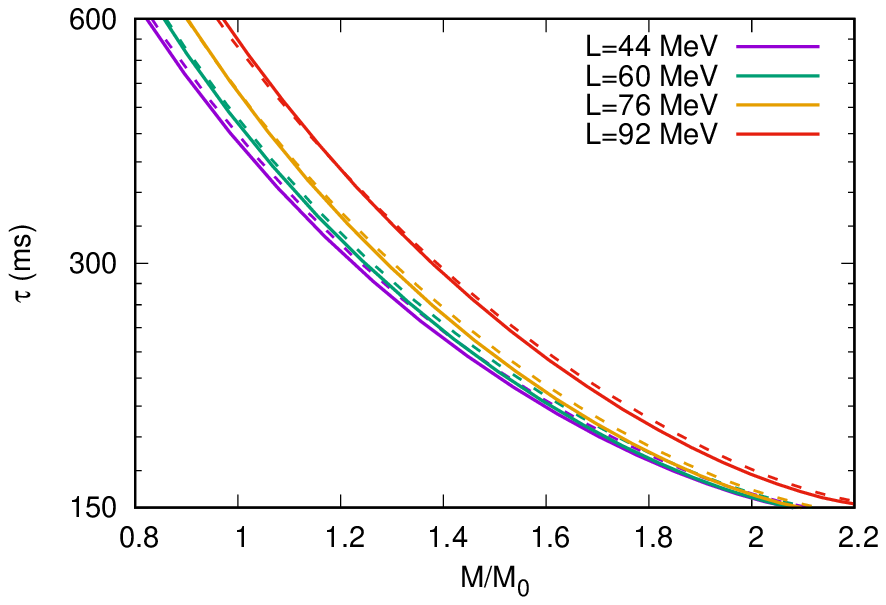} \\
\end{tabular}
\caption{\textbf{left}) Frequency of the fundamental mode (\textbf{right}) damping time. Solid lines indicate weakly-magnetized neutron stars, and dotted lines indicate strongly-magnetized ones\label{F3}} 
\end{figure*}

\begin{figure*}[ht]
\begin{tabular}{ccc}
\centering 
\includegraphics[scale=.58, angle=270]{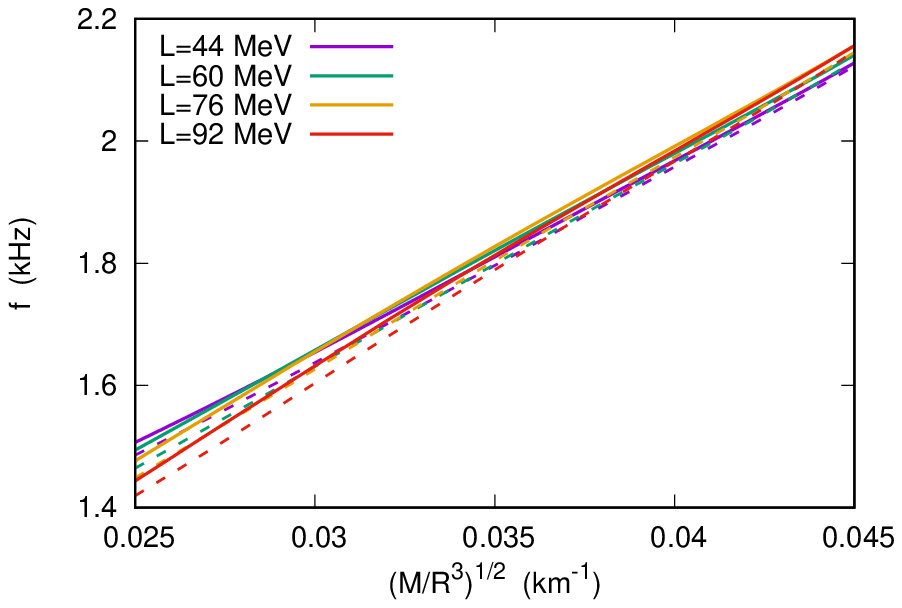}
\end{tabular}
\caption{The frequency of the fundamental mode is plotted in the upper
panel as a function of the square root of the average density for the
different EoSs. Solid lines indicate weakly-magnetized neutron stars, and dotted lines indicate strongly-magnetized ones\label{F4}} 
\end{figure*}

\begin{widetext}
\begin{center}
\begin{table}[ht]
\begin{center}
\scriptsize
\begin{tabular}{cc|cccccccccccccc}
\hline
  $L (\mathrm{MeV})$ &$B_0$ &$M~(M_\odot)$  & R (km) & $\Delta$R (km)  &$\Lambda$ & $\Delta\Lambda$  & $z$ & $\Delta z$  & $f$ (KHz) & $\Delta f$ (kHz) & $\tau$ (ms) & $\Delta \tau$ (ms) & $\epsilon_c$ (MeV/fm$^{3}$) \\
\toprule
 44 & B0 & 1.0 &  12.34 & - & 3026 & - & 0.146 & - &  1.592&  - & 424 & - & 330 \\
 44 &B1 & 1.0 & 12.28 & -0.06 & 2695 & -331 & 0.147 & +0.001 & 1.574 & -0.018 & 434
 & +10 & 323 \\
44 &B0 & 1.2 & 12.43 & -& 1200  & - & 0.183 & -  & 1.667 & - & 310 & - & 368 \\ 
44 &B1 & 1.2 & 12.40 &-0.03 & 1066 & -134 &0.183 & 0.000 & 1.648 & -0.019 & 317 & +7 & 359 \\
44 &B0 & 1.4 & 12.51 & - & 506  & - & 0.224  & - & 1.734 &  - & 243 & - & 406 \\
 44& B1 &1.4 & 12.52&  +0.01 & 475 & -31 & 0.224 & 0.000 & 1.714   & -0.020 & 258 &  +5 & 397  \\
 44&B0 &1.6 & 12.51 & - & 226 & - &0.269 & - & 1.801   &  - & 200 &  - & 456  \\
 44&B1 &1.6 & 12.55 & +0.04 & 209 & -17  &0.268 & -0.001 & 1.782  & -0.019 & 204 & +4 & 447  \\
 44&B0&2.0 & 12.35 & - & 45 & - & 0.388 & - & 1.954   &  - & 154 & - & 602  \\
 44&B1 &2.0 & 12.41 & +0.06 & 42 & -3  & 0.384 & -0.004 & 1.936   & -0.018 & 156& +2 & 589  \\
\hline
60 & B0 & 1.0 &  12.65 &- & 3011 & - & 0.143 & - & 1.558 &  - & 448 & - &  326 \\
 60 &B1 & 1.0 & 12.49 &-0.16 &2619 &-392 & 0.145 & +0.002 & 1.547 &  -0.011 & 455 & +7 & 319 \\
60 &B0 & 1.2 & 12.66 & - & 1154   & - & 0.178 & - & 1.638 &  - & 322 & - & 364 \\ 
60 &B1 & 1.2 & 12.57 &-0.09 & 1042 & -112 &0.180 & +0.002 & 1.624 &  -0.014 & 328 & +6 & 355 \\
60 &B0 & 1.4 & 12.71 & - & 501  & - &0.220  & - & 1.714 &  - & 248 & - & 405 \\
 60& B1 &1.4 & 12.66& -0.05 & 460 & -41 & 0.221 & +0.001 & 1.696   & -0.018 & 253 & +5 & 395  \\
 60&B0 &1.6 & 12.68 & - & 213 & - &0.267 & - & 1.788   &  - & 202 &  - & 455 \\
 60&B1 &1.6 & 12.66 &-0.02 & 201 & -12  &0.267 & 0.000 & 1.770   & -0.018 & 206 &  +4 & 446  \\
 60&B0&2.0 & 12.44 & - & 42 & - & 0.384 & - & 1.951  &  - & 154 & - & 605  \\
 60&B1 &2.0 & 12.47 & +0.03 & 40 & -2  & 0.382 & -0.002 & 1.933   & -0.018 & 156 & +2 & 593  \\
 \hline
76 & B0 & 1.0 &  13.04 & - & 3501 & - & 0.138 & - & 1.495 &  - & 488 & - & 314 \\
 76 &B1 & 1.0 & 12.87 &-0.17 &2807 & -694 & 0.140 & +0.002 & 1.495 &  0.000 & 488 & 0 & 306 \\
76 &B0 & 1.2 & 13.00 & - & 1307   & - & 0.173 & - & 1.589 & - & 342 & - & 355 \\ 
76 &B1 & 1.2 & 12.90 & -0.10 & 1099 & -208 &0.175 & +0.002 & 1.578 &  -0.011 & 347 & +5 & 346 \\
76 &B0 & 1.4 & 12.99 & - & 535  & - &0.213  & - & 1.677 & - & 260 & - & 399 \\
 76& B1 &1.4 & 12.93& -0.06 & 482 & -53 & 0.214 & +0.001 & 1.659   & -0.018 & 265 & +5 & 390  \\
 76&B0 &1.6 & 12.91 & - & 236 & - &0.258 & - & 1.759   &  - & 208 &  - & 451  \\
 76&B1 &1.6 & 12.89 & -0.02 & 208 & -18  &0.258 & 0.000 & 1.740   & -0.019 & 212 & +4 & 441  \\
 76&B0&2.0 & 12.58 & - & 43 & - & 0.376 & - & 1.939   &  - & 156 &  - & 606  \\
 76&B1 &2.0 & 12.61 & +0.03 & 41 & -2  & 0.374 & -0.002 & 1.919   &  -0.020 & 158 & +2 &594  \\
\hline
92 & B0 & 1.0 &  13.59 & - & 4111 & - & 0.131 & - & 1.400 &  - & 566 & - & 278 \\
 92 &B1 & 1.0 & 13.35 &-0.24 &3306 & -805 & 0.134 & +0.003 & 1.415 & +0.015 & 55 & -11 & 272 \\
92 &B0 & 1.2 & 13.56 & - & 1555  & - & 0.164 & - & 1.493 &  - & 388 & - & 319 \\ 
92 &B1 & 1.2 & 13.40  &-0.16 & 1307 & -248 &0.167 & +0.003 & 1.493 & 0.000 & 388 & 0 &  310 \\
92 &B0 & 1.4 & 13.49 & - & 640  & -  &0.202  & - & 1.582 &  - & 292 & - & 363 \\
 92& B1 &1.4 & 13.41 & -0.08 & 566 & -74 & 0.203 & +0.001& 1.571   & -0.011 & 295 & +3 & 355  \\
 92&B0 &1.6 & 13.41 & - & 287 & - &0.244 & - & 1.670   & - & 229 &  - & 416  \\
 92&B1 &1.6 & 13.37 & -0.04 & 251 & -36 &0.245 & +0.001 & 1.655   &  -0.015 & 233 & +4  &404 \\
 92&B0&2.0 & 13.05 & - & 55 & -  & 0.354 & - & 1.857   &  - & 165 &  - & 560  \\
 92&B1 &2.0 & 13.07 & +0.02 & 51 & -4  & 0.353 & -0.001 & 1.838   & -0.019 & 167 & +2 & 548   \\
\hline
\end{tabular}
\caption{Neutron stars' main properties for different values of $L$ with weak and strong magnetic field. } \label{T3}
\end{center}
\end{table}
\end{center}
\end{widetext}

The fundamental mode (\textit{f}-mode) corresponds to a class of non-radial oscillation modes in compact stars and is primarily determined by the global properties of the star, such as its mass and radius. These oscillations act as sources of gravitational waves, whose characteristic frequencies and damping times are highly sensitive to the EoS of the stellar interior. As a result, the \textit{f}-mode serves as a powerful diagnostic tool for probing the internal structure and composition of neutron stars.

We can see that, for low masses M$<1.4$M$_{\odot}$ and by increasing $L$, we obtain a decreasing of the gravitational wave frequencies, all below $f < 1.7$kHz. For masses higher than 1.4M$_{\odot}$ the effect is almost insignificant and all the frequencies converges to aproximately $f < 2.1$kHz. In the case of a strong magnetic field (dashed lines), for  masses  below M$<2.0 $M$_{\odot}$, the frequencies have a systematic shifting and the the difference is about 1.8 $\%$, but for masses above $2.0 $M$_{\odot}$ the magnetic field has not significant effect.
A similar behaviour is observed for the damping time. In fact, for massive stars, the damping time converges to 150 ms. We can conclude that the magnetic field does not have a significant effect on massive stars, but higher values of $L$ have a considerable effect.

Before finishing our analyses, we have investigated the universal relation between the frequency of the fundamental mode and the square root of the average density, $(M/R^3)^{1/2}$. It was pointed out in Ref.~\cite{Benhar2004} that in the Newtonian limit of the theory of stellar perturbations, the frequency scales as the
square root of the average density; and the gravitational wave frequency can be fitted by the following linear expression:

\begin{equation}
 f = a + b\cdot(M/R^3)^{1/2},   \label{sqd}
\end{equation}
where $a$ is given in kHz and $b$ in km $\times$ kHz.  These results are displayed in Fig.~\ref{F4} for masses above 0.65 $M_\odot$, and the calculated values of $a$ and $b$ are presented in Tab.~\ref{T4} altogether with some values that can be found in the literature. 

We notice that, as a general rule, as we increase the value of $L$ we reduce the value of $a$ and increase the value of $b$. A deviation of this rule is the value of $L = 60$ MeV. As happens with the dimensionless tidal parameter, this value of $L$ predicts the lower value of $b$ for all analyzed slopes. A strong magnetic field also acts to decrease the value of $a$ and increase the value of $b$.  We found a median value of $a = 0.611$ kHz and $b = 34.1$ km$\times$kHz for the weak magnetic field approach (very similar to the values found in ref.~\cite{Guha_2024APj}); and $a = 0.550$, $b = 35.3$km$\times$kHz for the strong magnetic field.

There are several promising avenues for measuring f-mode \cite{PhysRevD.65.063006,VásquezFlores_2014,PhysRevD.106.123002} characteristics in compact stars. For instance, third-generation gravitational wave detectors like the Einstein Telescope \cite{Punturo_2010} and Cosmic Explorer \cite{PhysRevLett.119.161101} are expected to achieve the sensitivity required for detecting f-mode signals from astrophysical sources \cite{PhysRevD.85.024030}.

\begin{center}
\begin{table}[ht]
\begin{center}
\begin{tabular}{cc|cc}
\hline
  $L$ (MeV) & $B_0$ & $a$ & $b$  \\
 \hline
 44 & B0 & 0.652 & 32.8 \\
 44 & B1 & 0.450 & 37.2 \\
 60 & B0 & 0.691 & 32.2 \\
 60 & B1 & 0.633 & 33.3 \\
 76 & B0 & 0.616 & 34.2 \\
 76 & B1 & 0.587 & 34.6 \\
 92 & B0 & 0.485 & 37.2 \\
 92 & B1 & 0.530 & 35.9 \\
 Mean & B0 & 0.611 & 34.1 \\
 Mean & B1 & 0.550 & 35.3 \\
 \hline
 Other works & Ref. & \\
 \hline 
 Anderson et al. & Ref.~\cite{Anderson1998} & 0.220 & 47.5 \\
  Pradhan et al. & Ref.~\cite{Prad2022} & 0.790  & 33.0   \\
 Benhar et al. & Ref.~\cite{Benhar2004} & 0.535 & 36.2   \\
 Guha Roy et al. & Ref.~\cite{Guha_2024APj} & 0.626 & 35.9 \\
 Chirenti et al. &Ref.~\cite{Chirenti:2015dda} & 0.332 & 44.0 \\
\hline 
\end{tabular}
 
\caption{ Values of fitting coefficients for Eq.~\ref{sqd} from different values of $L$ and magnetic field. Results presented in other works are also added.} 
\label{T4}
\end{center}
\end{table}
\end{center}

\section{Conclusions}

In this work, we studied the effects of a strong magnetic field for different values of the symmetry energy slope. 
Bearing in mind that our results were obtained with the chaotic field approximation,
the main results can be summarized as:

\begin{itemize}

    \item At very low density, larger slope values  produce soft EOS. This is reversed at low and moderate densities. In the high-density limit, all EOS become almost degenerate.

    \item Strongly magnetized EOS are soft at low densities but stiffer at moderate densities. The effect is more significant for higher values of the slope. 

    \item For a fixed mass, the neutron star radius grows with $L$. In the case of the magnetic field, we see that low mass stars bear 
    lower radii, but high mass stars actually present larger values of $R$. The mass for which the radii are the same for weakly and strongly magnetized neutron stars depends on $L$. For $L = 44$ MeV, this mass lies below 1.4 M$_\odot$. On the other hand, for $L = 92$ MeV, 
    $M~>1.8 M_\odot$.

    \item All models discussed in this work satisfy the constraints related to the  PSR J0740+6620 and the radius of the canonical star.

    \item The gravitational redshift is sensitive to the slope, but not to 
    the strong magnetic fields. In all cases, strong magnetic fields only produce absolute values of $\Delta z~<$ 0.005.

    \item 
     In contrast to the redshift, the tidal dimensionless parameter is very sensitive to the magnetic field. Within the weak magnetic field limit, we always 
    obtain $\Lambda_{1.4}~>500$. But, for the strong magnetic field, 
    $\Lambda~<500$ for three different values of $L$.
    In the lower mass limit, a strong magnetic field can produce an absolute value of $\Delta \Lambda~>$ 800. 

    \item We have seen that, in the case of the gravitational wave frequency, the magnetic field has a tiny effect on masses below $2.0 $M$_{\odot}$.

    \item Our results show that an increase in $L$, in the region of low masses, produces a decrease in the gravitational wave frequencies. In the case of large masses, $L$ does not produce any effect. 

    \item By analysing the universal relation of Eq.~\ref{sqd}, we see that an increase of the slope causes a decrase in $a$ but an increase in $b$. The same happens when we increase the strength of the magnetic field.

\end{itemize}

{\bf Funding:} 
This work is a part of the project INCT-FNA proc. No. 408419/2024-5. 
  It is also supported by Conselho Nacional de Desenvolvimento Cient\'ifico e Tecnol\'ogico (CNPq) under Grants No. 303490/2021-7 (D.P.M.), 305347/2024-1 (L.L.L.) and   304569/2022-4 (C.V.F.).

{\bf Author Contributions:} L.L.L. and C.V.F. developed the codes and obtained the data. All authors contributed to the development of the ideias, discussion of the results, and to the writing process.

{\bf Data Availability Statement:} Data will be made available upon reasonable request. Credits will be required. 

{\bf Conflicts of Interest:} No conflicts of interest have to be stated. 

\bibliographystyle{ieeetr}
\bibliography{aref}

\end{document}